\documentclass[letterpaper]{article}

\usepackage[T1]{fontenc}

\usepackage{geometry}
\usepackage{setspace}

\usepackage[style = chem-acs,articletitle=true]{biblatex}
\usepackage[colorlinks=true, linkcolor=blue, citecolor=blue, urlcolor=blue]{hyperref}
\usepackage{graphicx}
\usepackage{caption}
\usepackage{float}
\newfloat{scheme}{htbp}{los}
\floatname{scheme}{Scheme}
\floatname{chart}{Chart}
\newfloat{graph}{htbp}{loh}

\usepackage{chemformula} 
\usepackage[version = 4]{mhchem} 

\usepackage{authblk}

\author[1]{Riccardo Bertini}
\author[2]{Xueqiao Wang}
\author[3,4]{Sergey Slizovskiy}
\author[2]{Zhiren Zheng}
\author[1]{Julien Barrier}
\author[1]{Chiara Pizzo}
\author[5]{Robin Smeyers}
\author[1]{Krystian Nowakowski}
\author[1]{Hitesh Agarwal}
\author[1]{Alvaro Moreno}
\author[5]{Bert Jorissen}
\author[6]{Kenji Watanabe}
\author[7]{Takashi Taniguchi}
\author[5]{Milorad V. Milo\v{s}evi\'c}
\author[5]{Lucian Covaci}
\author[3]{Vladimir Fal'ko}
\author[2]{Pablo Jarillo-Herrero}
\author[1,8]{Roshan Krishna Kumar$^*$}
\author[1,9]{Frank H. L. Koppens$^\dagger$}

\affil[1]{ICFO - Institut de Ci\`encies Fot\`oniques, The Barcelona Institute of Science and Technology, Av. Carl Friedrich Gauss 3, 08860 Castelldefels (Barcelona), Spain}
\affil[2]{Department of Physics, Massachusetts Institute of Technology, Cambridge, MA, USA}
\affil[3]{National Graphene Institute, University of Manchester, M13 9PL Manchester, U.K.}
\affil[4]{Dept. of Physics \& Astronomy, University of Manchester, Manchester M13 9PL, U.K.}
\affil[5]{COMMIT, Department of Physics and NANOlight Center of Excellence, University of Antwerp, Groenenborgerlaan 171, 2020 Antwerp, Belgium}
\affil[6]{Research Center for Electronic and Optical Materials, National Institute for Materials Science, 1-1 Namiki, Tsukuba 305-0044, Japan}
\affil[7]{Research Center for Materials Nanoarchitectonics, National Institute for Materials Science, 1-1 Namiki, Tsukuba 305-0044, Japan}
\affil[8]{ICN2 - Institut Catal\`a de Nanoci\`encia i Nanotecnologia}
\affil[9]{ICREA - Instituci\'o Catalana de Recerca i Estudis Avan\c{c}ats, 08010 Barcelona, Spain}

\title{Bandwidth-Limited Critical Currents in Electrically Tunable Moiré Bands}
\date{*Email: roshan.krishna@icn2.cat, $\dagger$ Email: frank.koppens@icfo.eu }

\begin{document}

\maketitle

\begin{abstract}
  Moiré superlattices host narrow minibands whose bandwidth governs correlated and topological phases. Here, we demonstrate that the bandwidth also sets the critical current for the onset of out-of-equilibrium transport.  In bilayer graphene aligned to hexagonal boron nitride, we explore the high-current transport regime as we continuously flatten the valence miniband using an out-of-plane displacement field. We observe a significant reduction in the critical current, which is captured by a minimal analytical model and corresponds to the calculated narrowing of the miniband. Moreover, by comparing distinct moiré platforms, we show that the scaling between critical current and bandwidth is a universal feature of graphene superlattices. Our results reveal a direct link between miniband dispersion and high-current transport, and establish this regime as a fast and accessible electrical probe of bandwidth evolution.
\end{abstract}



\clearpage

\section{Introduction}

Two-dimensional (2D) moiré crystals have opened a new era in quantum materials, as they give access to tunable flat bands via structural control of superlattice potentials \cite{Yankowitz2012, Cao2018}. In many systems, 
the out-of-plane displacement field breaks additional crystal symmetries, providing electrical control over bandgaps and bandwidths \cite{zhangDirectObservationWidely2009a, Joucken2019, adakTunableBandwidthsGaps2020} and reshaping the quantum geometry of the bands \cite{hanOrbitalMultiferroicityPentalayer2023,Sinha2022}.
At its core, the interest of tunable narrow bands derives from their reduced carrier velocity: low velocity enhances the role of Coulomb interactions, promoting correlated quantum phases \cite{chenEvidenceGatetunableMott2019,hanCorrelatedInsulatorChern2024,Chen2019b}, and eventually stabilizing new magnetic orders \cite{Liu2020d, Chen2020f}. The same velocity scale should also be expected to govern charge transport, as flatter bands with slower carriers should naturally sustain smaller currents. 

\begin{figure*}[t]  
  \centering
  \includegraphics[width=\textwidth]{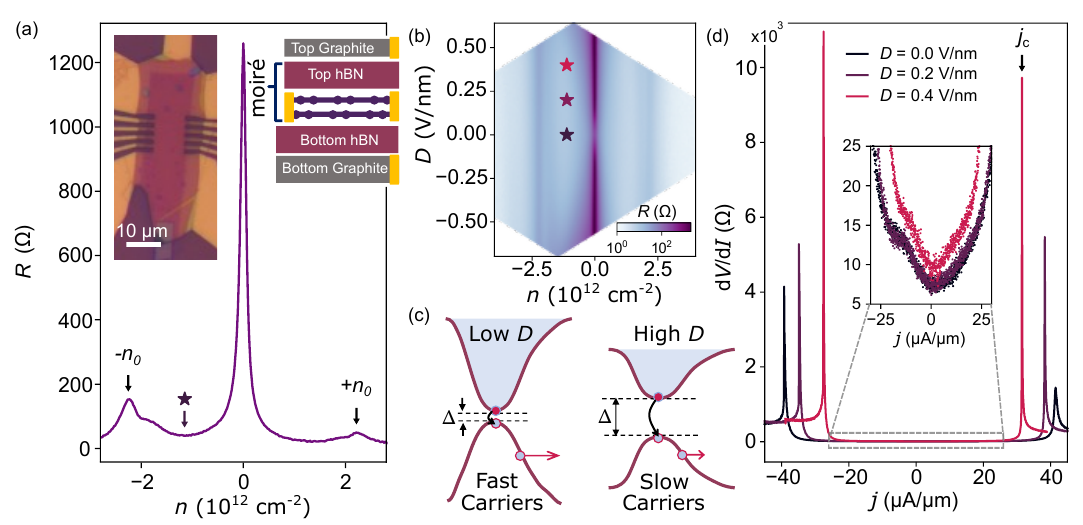}
  \caption{Low- and high-current transport regimes in BLG/hBN superlattices. (a) Four-probe resistance for device X3 at $D = 0$\,V/nm using 5\,nA/\textmu m current bias. Black arrows indicate the position of the secondary neutrality points (NPs) on the hole and electron side. Left inset: Optical image of the device. Right inset: Schematic of the dual-gated geometry. The estimate of twist angle from full filling peaks at secondary NPs gives $\theta \approx 0.13^{\circ}$. (b) Carrier density--displacement field map of the resistance in the low-current regime. The central indigo streak corresponds to the main NP feature and sidebands correspond to secondary NPs. Stars mark the configurations at which high-current measurements in panel (d) are performed. (c) Schematic representation of the band structure along high-symmetry lines and the main features influencing nonlinear transport: flatter bands result in lower carrier velocity and lower critical current at criticality, while larger gaps result in sharper $dV/dI$ peaks. (d) Differential resistance as a function of DC current density measured at $n/n_0 = -0.5$ and $D = 0$, 0.2, 0.4\,V/nm, respectively. The sharp peaks define the critical current at the transition between the Fermi liquid and the out-of-equilibrium regime. Inset: Zoom on the intermediate current regime showing deviations from the linear response due to heating effects but minor differences between different $D$ fields.}
  \label{fig:1}
\end{figure*}

Recent experiments have revealed striking critical-current phenomena in moiré superlattices under high in-plane bias \cite{Berdyugin2022,Tian2023, nowakowskiSinglephotonDetectionEnabled}. These observations point to a breakdown of equilibrium single-band transport and the onset of strongly nonlinear dynamics (Figure \ref{fig:1} (c)). However, the connection between this behavior and the underlying electronic dispersion has not been systematically established. Here, we demonstrate that the critical current in moiré superlattices obeys a universal scaling with electronic bandwidth, tunable by displacement field and robust across platforms. This identifies bandwidth as the quantity fundamentally limiting the maximum current sustainable by a moiré miniband, while also establishing nonlinear transport as a rapid electrical probe of changes in electronic bandwidth.

Most transport experiments are performed in the low-current regime, where the DC conductivity is mediated by carriers at the Fermi level, $E_F$. By contrast, high-current experiments drive electrons thermodynamically out-of-equilibrium. The strong in-plane electric field spreads the carriers across the energy band, opening new scattering channels and making transport sensitive to the electronic dispersion far from $E_F$, including bandwidths, heavy-mass states and gaps \cite{Balkan1993}. In this condition, the current-voltage ($I$--$V$) response becomes nonlinear. While many different nonlinear responses have been observed in 2D materials \cite{heNegativeDifferentialConductance2017,Berdyugin2022,Schmitt2023,Guo2025, Dong2025, andersenElectronphononInstabilityGraphene2019a}, a hallmark of metals under high fields is the appearance of a current-saturation regime, whose limiting current is set by the average group velocity of carriers. The electronic bandwidth $W$ and the size of the Brillouin zone $2\pi/d$ determine the intrinsic velocity scale
\begin{equation}
  \langle v\rangle \sim \frac{Wd}{2\pi \hbar},
  \label{eq:vmax_bandwidth}
\end{equation}
which represents an upper limit that can be reduced by scattering, phonons, and other dissipative mechanisms.
Moiré superlattices---with their small Brillouin zones and isolated, narrow minibands--- offer the possibility to reach current-saturating regimes at much smaller currents than in conventional metals. 
In this setting, critical-current behavior appears as a sensitive manifestation of the miniband structure.

In this work, we probe the high-current regime in Bernal-stacked bilayer graphene (BLG) aligned to hexagonal boron nitride (hBN), whose single-particle bands can be controlled \emph{in situ} using perpendicular electric fields \cite{Smeyers2023a,Shilov2024}. While tuning the displacement field, we observe a strong modification of the critical-current response that tracks the band narrowing and the opening of spectral gaps. We interpret the effect as a bandwidth-limited saturation arising from carriers slowing in increasingly flat minibands.

\section{Results}

\subsection{Current-induced criticalities in BLG/hBN superlattices}

\begin{figure*}[]  
  \centering
  \includegraphics[width=\textwidth]{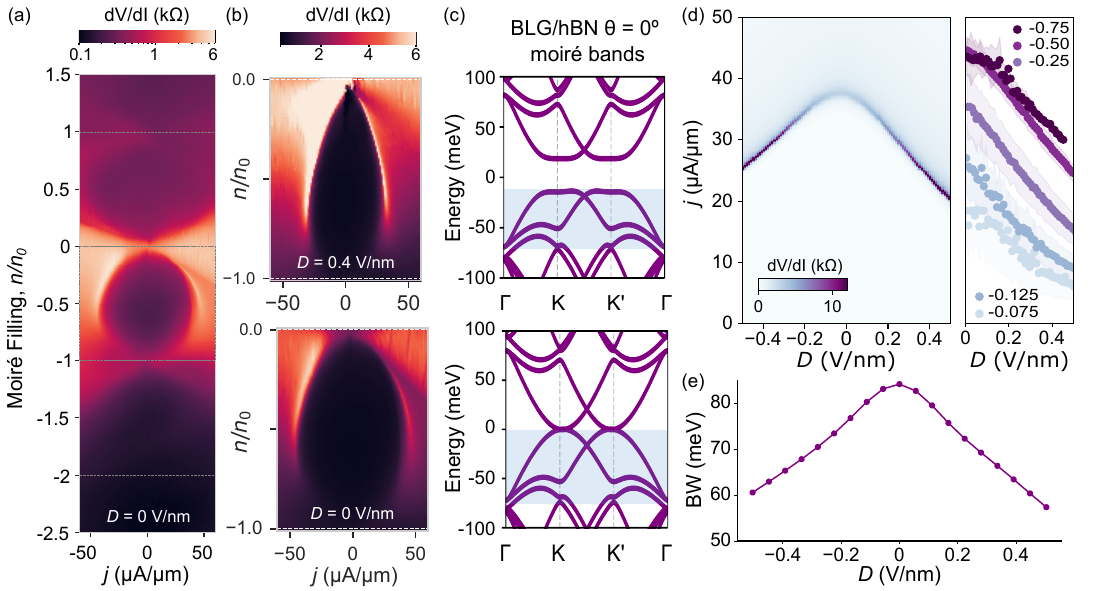}
  \caption{Tuning the critical regime with carrier density and displacement field. (a) Differential resistance as a function of bias current density and band filling at $D = 0$\,V/nm. Arc-like structures evidencing the evolution of peaks in $dV/dI$ emanate from the main NP and seemingly close at the secondary NPs. A strong asymmetry in the sharpness of the criticality is present between hole and electron filling. Additional features are also visible in the remote bands beyond $|n/n_0|>1$. (b) Zoom of $dV/dI$ maps in the first valence band for $D = 0$, 0.4\,V/nm (bottom to top). The criticality arcs are visible starting from the main NP and disappearing after $n/n_0 \approx -0.5$, shrinking progressively as $|D|$ is increased. (c) Band-structure calculations for BLG/hBN superlattices at $n/n_0 = -0.5$ corresponding to the same $D$ fields as in panel (b) calculated with a real-space method. A band gap opens as $D$ is increased, and the band edge becomes flatter. The shaded blue line indicates the first valence band which flattens at high $|D|$. (d) Color plot of differential resistance $dV/dI$ as a function of current density and displacement field at fixed moir\'e filling $n/n_0 = -0.5$ in BLG X3. Dark purple streaks show the critical current density for positive and negative displacement field, reducing monotonically as $|D|$ is increased. Right panel: Critical current scaling with $|D|$ extracted from measurements as in panel (d) at different moir\'e fillings, in BLG 39\_I. The shaded area corresponds to half the difference between critical currents measured at positive and negative bias. (e) Bandwidth (BW) of the first valence band calculated as the difference between the maximum and the minimum energy of the valence band, as a function of $D$ field extracted from tight-binding calculations for BLG/hBN at $n/n_0 = -0.5$.}
  \label{fig:2}
\end{figure*}

We studied four BLG/hBN superlattices with moiré periods $d \sim 13\,$nm. In all devices, the top hBN layer was aligned with the BLG, while the bottom hBN was intentionally misaligned to avoid competing moiré patterns. Thin graphite gates were used on both sides of the heterostructures, allowing for independent tuning of charge carrier density and displacement fields while ensuring high electronic quality \cite{zibrovTunableInteractingComposite2017}. Devices were shaped into Hall bar geometries using standard nanolithography techniques, as shown in the inset of  Figure \ref{fig:1}(a). Fabrication details are detailed in the Supporting Information Section~1. All the measurements were conducted at cryogenic temperatures ($\sim 10$\,K) and data were corrected for self-gating effects as explained in Methods. In the main text, we focus on a single representative device; the results were consistent and reproducible across all devices (see Supporting Information Section 4). 

We first characterized our device in the linear regime, corresponding to low current biases ($\sim 2-10$\,nA/\textmu m). The measured four-probe longitudinal resistance $R$ was measured as a function of carrier density $n$ at fixed displacement field $D = 0$\,V/nm (see Methods). The resistance features a peak at the charge neutrality point (NP) and two peaks at densities $\pm n_0 \approx 2.4\cdot 10^{12}$ cm$^{-2}$ (see Figure~\ref{fig:1}(a)), corresponding to secondary NPs upon full filling of the moiré Brillouin zone~\cite{Bocarsly2024,Shilov2024}. Notably, the resistance peaks at the secondary NPs are much smaller than the main NP, characteristic of a low density of states rather than pronounced spectral gaps often found at the band edges of moiré superlattices \cite{caoSuperlatticeInducedInsulatingStates2016}. By independently varying top and bottom gate voltages, we tune both $n$ and $D$ in the device. A dual-gate map acquired in this way is reported in Figure~\ref{fig:1}(b). The application of out-of-plane displacement fields opens a gap at the main NP, as evidenced by a large increase in resistance (vertical indigo streak in Figure~\ref{fig:1}(b)) and activation behavior observed in temperature-dependent measurements (see Supporting Information Section 3), while the secondary NPs remain weakly influenced, leading to an overall flattening of the first minibands.

To study the high-current electrical response of this superlattice, we fixed the carrier density to half filling of the first valence miniband ($n/n_0 = -0.5$, marked with a star in Figure~\ref{fig:1}(a)) and measured the differential resistance $dV/dI$ as a function of DC current bias $j$. Figure~\ref{fig:1}(d) plots the high-current response up to $j\sim 45$\,\textmu A/\textmu m for three displacement fields (corresponding to the colored stars in Figure~\ref{fig:1}(b)). For each curve, the measured $dV/dI$ exhibit a sharp peak at a critical current $j_{\rm c}$. At the transition, the system shows a strong current saturation, marked by a pronounced feature in $dV/dI$ and flattening of the $I$--$V$ curve (see Supporting Information Section 4). This condition, known as out-of-equilibrium criticality~\cite{Berdyugin2022}, is currently understood to mark the transition between a velocity-limited metallic state~\cite{Tian2023,nowakowskiSinglephotonDetectionEnabled} and a Schwinger/Zener regime~\cite{Schmitt2023,Allor2008} where electron-hole pairs are created by the strong in-plane electric field, allowing current to grow beyond the single-band limit.

Notably, $dV/dI$ is dependent on current bias even for small values far before $j_{\rm c}$, possibly due to current-induced Joule heating~\cite{andersenElectronphononInstabilityGraphene2019a,He2021a}. However, at these biases the response remains little influenced by $D$, exhibiting less than 10\% variations in $dV/dI$ (Figure~\ref{fig:1}(d), inset) which correlates with mobility variations (see Supporting Information Section 2). In contrast, the response close to $j_{\rm c}$ is strongly modified by the displacement field. As $|D|$ increases, the critical current decreases by almost 50\% and the transition to the out-of-equilibrium state sharpens, as evidenced by higher peaks in $dV/dI$.

\subsection{Tuning nonlinearities in moiré superlattices with perpendicular electric fields}

Mapping the full density and displacement field dependence of the high-current response reveals how miniband changes influence the nonlinear transport. Figure~\ref{fig:2}(a) presents $dV/dI$ as a function of $j$ and moiré filling at fixed $D = 0$\,V/nm. As the carrier density is varied, sharp $dV/dI$ peaks form arc-like structures emanating from each NP and closing at the subsequent NPs, with similar qualitative responses in different bands. The $dV/dI$ peaks are sharper in the first valence band compared to the conduction and remote bands, revealing a strong electron--hole asymmetry even away from the secondary NPs. For this reason, we now focus on the first valence miniband, where the critical behavior is most pronounced between $-1 < n/n_0 < 0$, fading out approximately near the van Hove singularity~\cite{Shilov2024}, where the large density of states and indirect overlap with the second valence band, see Figure \ref{fig:2}c,  likely broadens the transition beyond detection.

We repeated the previous measurements at large $|D|$. Figure~\ref{fig:2}(b) plots zoomed maps of $dV/dI$ for moiré filling in the first valence band at $D = 0$ (bottom panel) and 0.4\,V/nm (top panel). At high $|D|$, the arc traced by $j_{\rm c}$ narrows to lower current values at all fillings $n/n_0$.  

For completeness, we show the entire dependence of $dV/dI$ on $j$ and $D$ at $n/n_0 = -0.5$ in Figure~\ref{fig:2}(d). The critical current $j_{\rm c}$, marked by slanted indigo streaks, decreases monotonically with $|D|$, while the peak height increases markedly (see Supporting Information Section 6 for extended data). Importantly, the reduction of $j_{\rm c}$ with $|D|$ is observed at all densities, as reported in the right panel of Figure~\ref{fig:2}(d), highlighting that the reduction of $j_{\rm c}$ is a band-wide effect, not specific to a particular filling. 

To quantify the effect of $D$ on the electronic bands, we perform large-scale tight-binding calculations following the approach of Ref.~\cite{Smeyers2023a}. Our calculations, reported in Figure~\ref{fig:2}(c) for $n/n_0 = -0.5$ and $D = 0$, 0.4\,V/nm, show that higher $|D|$ opens a bandgap at the main neutrality points leading to a monotonic reduction in bandwidth, as shown in Figure~\ref{fig:2}(e).
Complementary experimental evidence of band flattening is also provided by the enhancement of $dR/dT$ with $|D|$ (see Supporting Information Section~5~), consistent with reduced carrier velocity and an increased density of states in flatter bands.

\section{Discussion}

\subsection{Changes in the critical current track the bandwidth narrowing}

\textcolor{black}{Our experiments demonstrate that the out-of-plane displacement field strongly modifies the high-current transport of BLG/hBN superlattices. Three key observations are important for the discussion. First, the critical current $j_{\rm c}$ (Figure \ref{fig:2}(d)) follows the same evolution with $D$ as the miniband width (Figure \ref{fig:2}(e)). Second, while the critical current is highly sensitive to $D$, the low-current response is little influenced, as shown in the inset of Figure \ref{fig:1}d}. Third, the dependence $j_{\rm c}(D)$ is reproduced consistently across multiple devices, despite large variations in their low-current mobility (see Supporting Information Section 2). Together, these points suggest that high-current transport is governed by an energy scale of the superlattice which is largely independent of device-specific disorder or scattering. Our results identify this scale with the miniband width. 

\begin{figure}[]   
  \centering
  \includegraphics[width=0.5\linewidth]{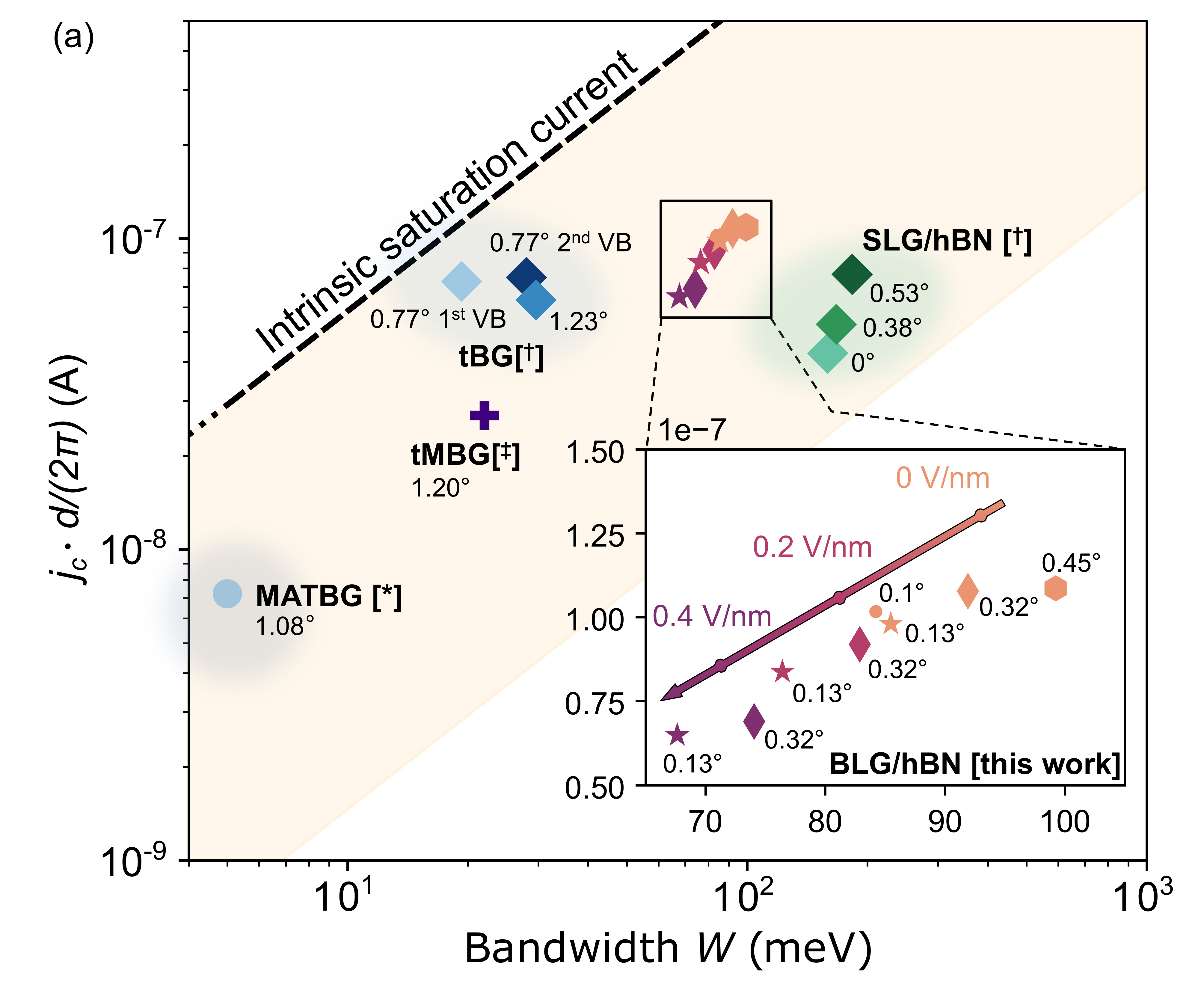}
  \caption{(a) Critical current vs bandwidth for different moiré superlattices. The experimental critical currents are extracted from $j_{\rm c}$ at $n/n_0 = -0.5$ for the available data in the literature: $\dagger$ from Berdyugin \textit{et al.}~\cite{Berdyugin2022}, $\ddagger$ Barrier~\cite{BarrierThesis}, $*$ Tian \textit{et al.}~\cite{Tian2023}. The currents are normalized by BZ size $2\pi/d$ to isolate the bandwidth dependence according to Eq.~\ref{eq:jmax_model}. Bandwidths for all the samples are calculated using the continuum model. The dashed black line represents the maximum current limit according to Eq.~\ref{eq:jmax_model}. Inset: Same as main panel, but for the different BLG/hBN superlattices measured in this work, different colors correspond to the same device measured at different displacement fields (orange: $D = 0$\,V/nm, pink: $D = 0.2$\,V/nm, purple: $D = 0.4$\,V/nm).}
  \label{fig:3}
\end{figure}

To verify this trend further, in the inset of  Figure~\ref{fig:3} we compile $j_{\rm c}$ as a function of miniband width for the four devices measured in this work, at the same filling ($n/n_0 = -0.5$) of the first valence miniband. Each marker type corresponds to a different device (identified by the BLG-hBN twist angle extracted from the position of secondary NPs) while different $D$ fields are identified with different colors. Comparing these samples reveals a clear trend: $j_{\rm c}$ scales with the bandwidth, decreasing for narrower bands at high $|D|$, consistent with our results in Figures~\ref{fig:1} and \ref{fig:2}. Changes in $j_{\rm c}$ among devices are also consistent with twist angle/moiré period variations, which sets the bandwidth as expected from band-structure calculations (see Supporting Information). 
We further compiled our data together with data from other graphene-based moiré systems reported in the literature~\cite{Berdyugin2022,Tian2023,BarrierThesis} (Figure~\ref{fig:3}). When grouping devices by superlattice type (TBG, BLG/hBN, etc.), the same tendency emerges: the critical current is proportional to the bandwidth. Although the absolute current values vary with superlattice type, likely reflecting differences in scattering mechanisms in different crystals, the bandwidth dependence itself is robust. Among the systems compiled, SLG/hBN exhibits the largest offset from the global trend line, although the scaling of $j_{\rm c}$ with bandwidth is preserved. This offset likely reflects a different scattering-related prefactor for this superlattice type, combined with a sizable spread in reported bandwidth values for SLG/hBN---ranging from ${\sim}100$\,meV \cite{moonElectronicPropertiesGraphene2014} to ${\sim}200$\,meV \cite{Tomadin2019PlasmonsMoire} depending on the tight-binding parametrization---which introduces additional horizontal uncertainty in the comparison. Additionally, across essentially all of the devices examined, smaller bandwidths lead to smaller critical currents, highlighting a generic link between high-current transport and the underlying electronic dispersion.

\begin{figure*}[]  
  \centering
  \includegraphics[width=\textwidth]{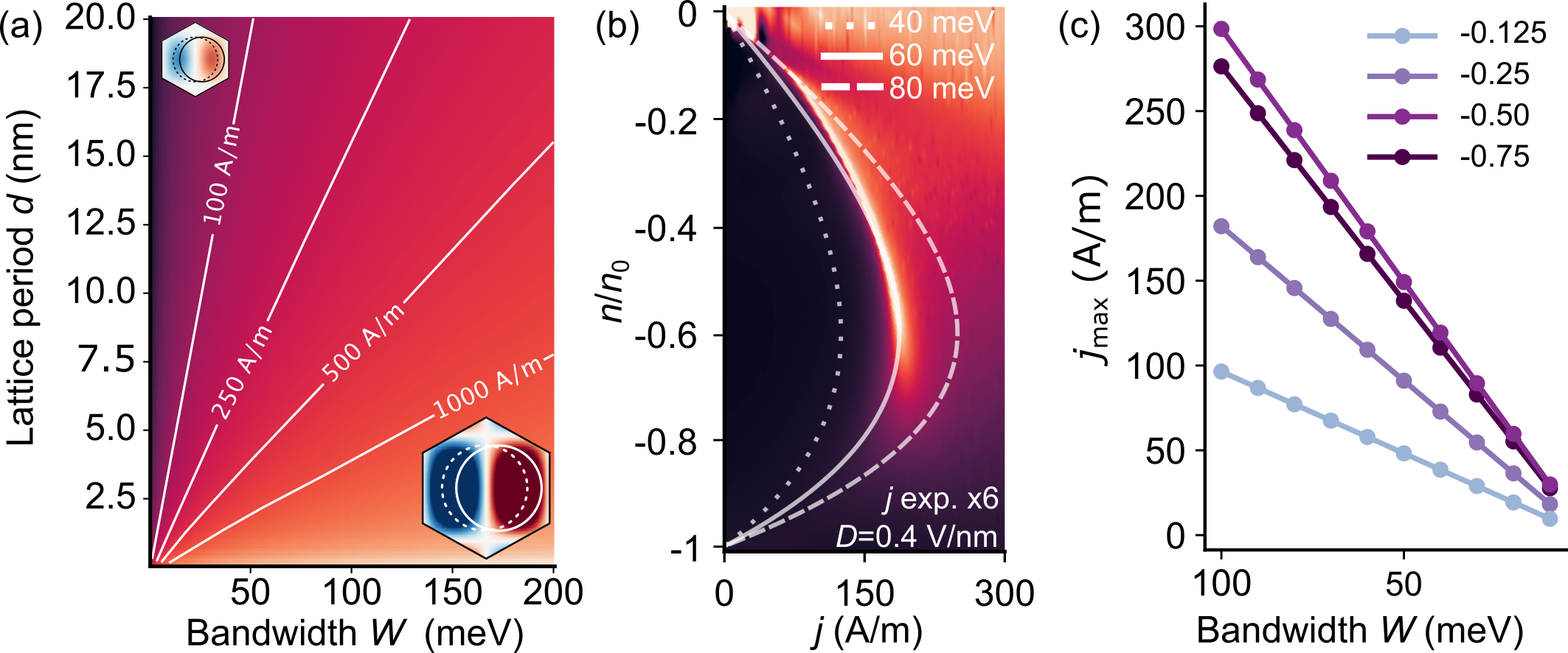}
  \caption{(a) Schematic illustration of the maximum current in a miniband, which grows with bandwidth $W$ and decreases with $d$. The insets show the BZ corresponding to small and large moiré period, colored with the band velocity $v_{k_x}$ and the shifted fermi surface along the field direction. (b) The different lines represent the density dependence of the maximum current obtained from the model, displaying arc-like structures similar to the experimental $dV/dI$ peaks. The colormap shows the same data as in the lower panel in Figure \ref{fig:2}b, with the experimental current multiplied by a factor 6 to show the qualitative agreement with the model for $W$ = 60 meV. (c) Bandwidth dependence of the critical current at fixed filling, showing an approximately linear decrease with decreasing bandwidth, in good agreement with the experimental trends in Figure~\ref{fig:2}(d) at different densities.}
  \label{fig:4}
\end{figure*}

\subsection{Boltzmann analysis of high-current transport} 

The connection between lower currents and flatter bands is contained in the general expression
\begin{equation}
  j = e \int \frac{d^2\mathbf{k}}{(2\pi)^2}\, f_{\mathbf{k}}(\mathbf{E})\, v_{\mathbf{k}},
  \label{eq:j_general}
\end{equation}
where $f_{\mathbf{k}}(\mathbf{E})$ is the Fermi distribution under electric field $\mathbf{E}$ and $v_{\mathbf{k}} = \frac{1}{\hbar}\nabla_{\mathbf{k}}\epsilon_{\mathbf{k}}$ is the band velocity, which decreases as the band dispersion $\epsilon_{\mathbf{k}}$ becomes flatter. In the low-$T$, equilibrium limit where $f_{\mathbf{k}}$ is sharply peaked at the Fermi surface, Eq.~(\ref{eq:j_general}) reduces to the usual Drude formula:
\begin{equation}
    j = nev_{\rm d},
    \label{eq:j_nevd}
\end{equation} where $v_{\rm d}$ is the average drift velocity. 

To reconcile the Drude picture with the notion that only carriers near the Fermi level contribute to transport at low currents, Eq. \ref{eq:j_nevd} should be interpreted as describing a current carried by a fraction of carriers, $n v_{\rm d}/v_{\rm F}$ each moving at the Fermi velocity $v_{\rm F}$. However at high currents, when carriers populate different states across the band, the Drude equation breaks down. To address this regime, we develop a minimal model that highlights the bandwidth dependence of $j_{\rm c}$ and provides insight into its behavior. 

In previous works, $j_{\rm c}$ was interpreted as a current-saturating process within a single miniband~\cite{Berdyugin2022,Tian2023,nowakowskiSinglephotonDetectionEnabled}. Building on this picture, we build a minimal model for the maximum current that a single band can sustain. Solving the Boltzmann transport equation analytically for a $C_3$-symmetric sinusoidal band in the relaxation-time approximation, we find that this maximum current can be written as
\begin{equation}
  j_{\max} = g_{\rm v,s}\, C\!\Big(\frac{n}{n_0}\Big)\, j_0, \qquad j_0 = \frac{8\sqrt{6}}{27}\,\frac{e}{\hbar}\,\frac{W}{d},
  \label{eq:jmax_model}
\end{equation}
where $g_{\rm v,s} = 4$ is the band degeneracy and the numerical prefactor of $j_0$ combines the geometry of the $C_3$-symmetric BZ with the shape function of the current--field relation evaluated at the critical field (see Supporting Information Section~8). A central feature of this result is the dimensionless coefficient $C(n/n_0)$, which depends only on the band filling and encodes the redistribution of carriers in momentum space that maximizes the current; at optimal filling it reaches $C \approx 0.057$. Equation~\ref{eq:jmax_model} thus generalizes Eq.~\ref{eq:j_nevd} to out-of-equilibrium electrons, while recovering also the saturation-velocity scaling reported for 1D superlattices (Eq.~\ref{eq:vmax_bandwidth}; see also \cite{Esaki1970, nazareneBLOCHOSCILLATIONSLN}).

Remarkably, the model reproduces both the density dependence--the arc-like closing of $j_{\rm c}$ between neutrality points (Figure~\ref{fig:4}(b)), and the linear scaling $j_{\rm c} \propto W$ (Figure~\ref{fig:4}(c)) without any fitting to the functional form, indicating that the shape of the critical-current response is set by the band structure alone. Importantly, it shows that the maximum current is set by the combination of bandwidth ($W$) and moiré period ($d$), as sketched in Figure~\ref{fig:4}(a). This can be understood intuitively by recalling that current always tracks the distribution of carriers in the superlattice minibands (Figure~\ref{fig:4}(a), insets). As such, it depends not only on their individual group velocity ($v_{\mathbf{k}}\propto Wd$), but also on the number of available states they can populate in the direction of the applied field, which scales as the area of the BZ ($n \propto 1/d^2$).

 However, in all cases, the measured $j_{\rm c}$ is systematically lower than the intrinsic limit (dashed line in Figure~\ref{fig:3}) by an approximately constant prefactor ($\sim 6\times$). This discrepancy suggests the presence of additional scattering mechanisms that preserve the underlying bandwidth scaling. Among these, acoustic phonon scattering is a natural candidate. However, although scattering could smear the distribution, it is unlikely to saturate the current. This could only happen in special circumstances related to phonon amplification where the critical current would be set by the sound velocity and, therefore, be independent of $D$. This is inconsistent with the $D$-dependent critical velocity measured in our experiments (see Supporting Information Section 4).

A comparison with other graphene superlattice devices reported in the literature reveals the same behavior: the arc structure~\cite{Berdyugin2022} and bandwidth scaling are universally present, pointing towards a more general connection between the critical current and bandwidth. Understanding the origin of the constant prefactor between $j_{\rm c}$ and $j_{\rm max}$, plausibly related to electronic scattering, remains an open question for further investigation.

\section{Conclusions}

In conclusion, we have shown that
the critical-current behavior in BLG/hBN closely tracks the evolution of the bandwidth, and that this scaling can be probed electrically by tuning the displacement field.  A minimal analytical model captures the functional scaling of $j_{\rm c}$ with both carrier density and bandwidth, confirming that the critical current is governed by the miniband dispersion up to a constant prefactor. Our analysis evidences that the same scaling law is valid across different moiré platforms, establishing a new universal boundary to the high-current regime of miniband transport.

Beyond offering new insights on the nature of out-of-equilibrium criticalities in moiré superlattices, our results establish high-current transport as a simple \emph{in situ} probe of miniband evolution. This is particularly important because bandwidth, while central to the physics of correlated moiré materials, remains difficult to measure directly and is typically accessible only through the most sensitive spectroscopic techniques \cite{Inbar2023b,Utama2021a,Xie2019b,Pasquale2023,Bocarsly2024} or through transport in special device architectures \cite{jiangDirectProbingEnergy2025} or miniband structures \cite{adakTunableBandwidthsGaps2020}. By demonstrating that the high-current response reflects the dispersion of the full miniband, we provide a fast, device-level electrical diagnostic that can be integrated directly into standard quantum-transport measurements. This capability may prove especially valuable for elucidating the parent bands of fractional Chern insulators~\cite{luFractionalQuantumAnomalous2024} and superconductors~\cite{hanSignaturesChiralSuperconductivity2025}, whose bandwidths are strongly tunable by displacement field and may be substantially renormalized by carrier density~\cite{choiInteractiondrivenBandFlattening2021}.

\section*{Acknowledgments}
We thank Prof. Leonid Levitov, Prof. Shahal Ilani, Prof. Ady Stern, Yael Rich, Atri Dutta, Shimon Uri for fruitful discussions on the high-bias measurements. We acknowledge Dr. Zoe Velluire-Pellat, Dr. Subhajit Sinha and Lorenzo Cavicchi for useful insight during manuscript and figures preparation. \newline

AI-use disclosure: During the preparation of this work, the authors used Axiomatic AI tools to assist with verification and validation checks of the high-bias analytical model. The tools were used solely to support internal consistency checks of derivations. All results, interpretations, and conclusions were independently verified by the authors, who take full responsibility for the content of the manuscript.\newline

R.B. acknowledges funding from the European Union’s Horizon 2020 research and innovation program under the Marie Sk\l{}odowska-Curie grant agreement No.~847517. K.W. and T.T. acknowledge support from the JSPS KAKENHI (Grant Numbers 21H05233 and 23H02052), the CREST (JPMJCR24A5), JST and World Premier International Research Center Initiative (WPI), MEXT, Japan. H.A. acknowledges funding from the European Union’s Horizon 2020 research and innovation program under Marie Sk\l{}odowska-Curie grant agreement No.~665884. R.S., B.J., M.V.M. and L.C. acknowledge support from Research Foundation–Flanders (FWO) through the FWO-FNRS EOS/ShapeMe, FWO/G0A5921N and FWO/11E5821N projects and computational resources provided by the VSC (Flemish Supercomputer Center). J.B. acknowledges support from the European Union’s Horizon Europe program under grant agreement 101105218. This material is based upon work supported by the Air Force Office of Scientific Research under award number FA8655-23-1-7047. A. M. acknowledges a Severo Ochoa Excellence Predoctoral Fellowship (CEX2019-000910-S)

\section*{Author contribution}

R.B., J.B., R.K.K. and F.H.L.K. conceived the experiments. R.B., J.B., C.P. performed the measurements and analyzed the data. K.N. and A.M. provided additional support to the experimental part. X.W. and Z.Z. fabricated devices BLG 39\_I, BLG 39\_III, BLG 39\_IV and BLG X3 supported by P.J.-H. H.A. fabricated device BLG A. S.S. performed high-bias calculations and self-gating calculations and provided valuable theoretical support for understanding high-current transport together with V.F. S.S., R.B., R.K.K. and F.H.L.K developed the minimal analytical model. R.S. and B.J. performed real-space calculations of BLG/hBN superlattices supported by M.V.M. and L.C. R.B. performed band-structure calculations using the Python package Pyband-structure~2.1. T.T. and K.W. provided the high-quality hBN crystals. R.B., R.K.K., J.B. and F.H.L.K. wrote the manuscript with input from all authors.

\section{Conflict of Interest Statement}
The authors declare that there are no conflicts of interest related to this manuscript.
\section{Data Availability Statement}
All data presented in the main text are publicly available in a repository on Zenodo: https://doi.org/10.5281/zenodo.19005153. All other relevant data and additional supporting information are available upon reasonable request to the corresponding authors.

\printbibliography

\end{document}